\documentclass[reprint, two column, floatfix, prl, aps, showkeys]{revtex4-1}

\usepackage{amssymb}
\usepackage{amsmath}
\usepackage{graphicx}
\usepackage{float}
\usepackage{dcolumn}
\usepackage{bm}
\usepackage{color}
\usepackage{ulem}
\usepackage{float}
\usepackage{textcomp}
\restylefloat{table}
\usepackage{verbatim}
\usepackage{epstopdf}
\usepackage[mediumspace,mediumqspace,squaren]{SIunits}

\begin{document}

\title{Precise experimental test of the Luttinger theorem and particle-hole symmetry for a strongly correlated fermionic system}

\date{\today}

\author{Md.\ Shafayat Hossain}
\author{M. A.\ Mueed}
\author{M. K.\ Ma}
\author{K. A.\ Villegas Rosales}
\author{Y. J.\ Chung}
\author{L. N.\ Pfeiffer} 
\author{K. W.\ West}
\author{K. W.\ Baldwin}
\author{M.\ Shayegan}
\affiliation{Department of Electrical Engineering, Princeton University, Princeton, New Jersey 08544, USA}

\begin{abstract}

A fundamental concept in physics is the Fermi surface, the constant-energy surface in momentum space encompassing all the occupied quantum states at absolute zero temperature. In 1960, Luttinger postulated that the area enclosed by the Fermi surface should remain unaffected even when electron-electron interaction is turned on, so long as the interaction does not cause a phase transition. Understanding what determines the Fermi surface size is a crucial and yet unsolved problem in strongly interacting systems such as high-$T_\mathrm{c}$ superconductors.  Here we present a precise test of the Luttinger theorem for a two-dimensional Fermi liquid system where the exotic quasi-particles themselves emerge from the strong interaction, namely for the Fermi sea of composite fermions (CFs). Via direct, geometric resonance measurements of the CFs' Fermi wavevector down to very low electron densities, we show that the Luttinger theorem is obeyed over a significant range of interaction strengths, in the sense that the Fermi sea area is determined by the density of the \textit{minority carriers} in the lowest Landau level. Our data also address the ongoing debates on whether or not CFs obey particle-hole symmetry, and if they are Dirac particles. We find that particle-hole symmetry is obeyed, but the measured Fermi sea area differs quantitatively from that predicted by the Dirac model for CFs. 

\end{abstract} 

\maketitle 

\begin{figure*}[t!]
\includegraphics[width=.995\textwidth]{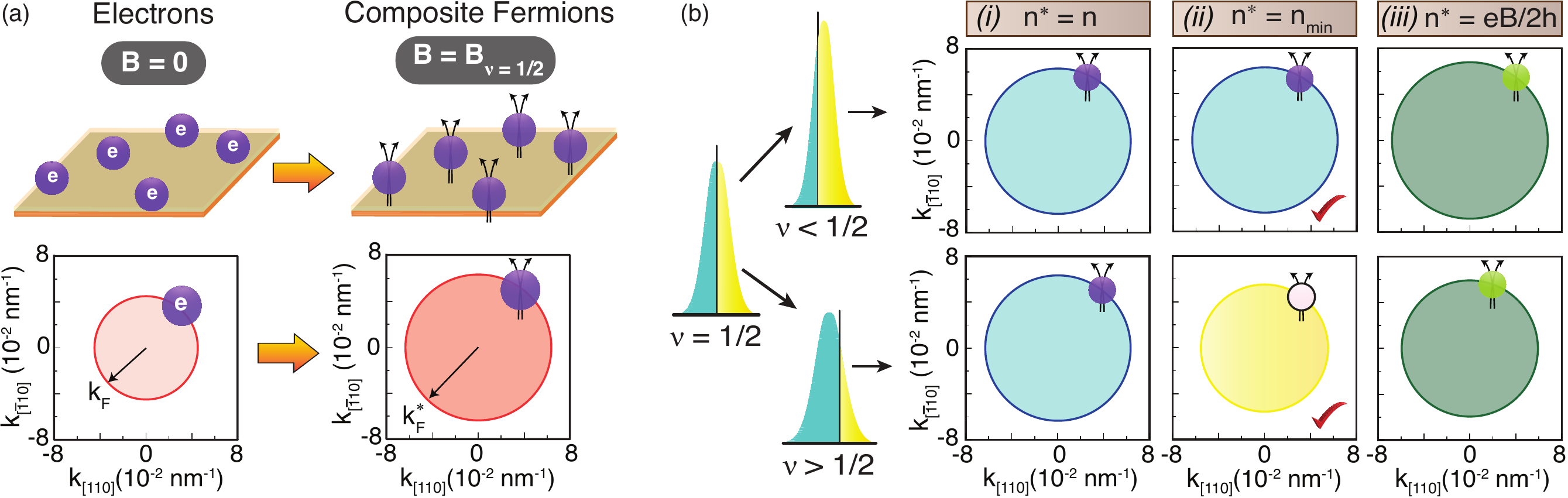}
\caption{\label{fig:Fig1} (a) Top panels: Electrons (at $B=0$) and CFs (at $\nu=1/2$) in real space. Bottom panels: Fermi seas of electrons and CFs at $n=3.20\times 10^{10}$ cm$^{-2}$, and their respective Fermi wavevectors ($k_\mathrm{F}$ and $ k_\mathrm{F}^*$), in reciprocal space. (b) The lowest LL at $\nu=1/2$ and its evolution away from $\nu=1/2$ at a fixed density ($n=3.20\times 10^{10}$ cm$^{-2}$) and varying magnetic field. The shaded regions denote the occupation of the lowest LL by electrons (blue) and holes (yellow). Our experimental data show that, out of the cases \textit{(i)} to \textit{(iii)} as described in the text, the CF Fermi sea area is determined by the density of \textit{minority carriers} [case \textit{(ii)}], namely by electrons ($n_\mathrm{min} =n$) for $\nu < 1/2$ and by holes [$n_\mathrm{min} = n(1-\nu)/\nu$] for $\nu > 1/2$, regardless of the interaction.}
\end{figure*} 

Composite fermions (CFs) are emergent quasi-particles of a strongly interacting, two-dimensional electron system (2DES) at high perpendicular magnetic fields when the electrons' kinetic energy is quenched into a set of quantized energy levels, the so-called Landau levels (LLs) \cite{Jain.PRL.1989, Halperin.PRB.1993, Jain.2007}. In the lowest LL, the electrons have no kinetic energy and the system is a prime example of a flat band system where interaction dominates the physics. When the lowest LL is half-filled, i.e., the LL filling factor ($\nu$) equals $1/2$, the interacting electrons each pair with an even number of flux quanta and form flux-electron CFs that condense into a metallic phase with a well-defined Fermi sea [Fig. 1(a)] \cite{Jain.PRL.1989, Jain.2007, Halperin.PRB.1993}. Thus the CF Fermi sea is born out of strong interaction within a flatband system, and provides an ideal platform to test the Luttinger theorem \cite{Luttinger.1960}, a major theorem in many-body physics which postulates that the Fermi sea and its area should be resilient against interaction. Here we investigate the validity of the Luttinger theorem and its link \cite{Heath.preprint, Seki.2017} to particle-hole symmetry [Fig. 1(b)], in a nearly half-filled, flatband system of interacting CFs. 

According to the CF theory \cite{ Jain.PRL.1989, Halperin.PRB.1993, Jain.2007}, the flux-electron CFs ignore the large, external magnetic field ($B$) and only experience an \textit{effective} magnetic field $B^{*} = B-B_{\nu=1/2}$, where $B_{\nu=1/2}$ is the field at $\nu=1/2$, $\nu=hn/eB$ is the LL filling factor, and $n$ is the 2DES density. Near $\nu=1/2$ CFs execute cyclotron motion in a small $B^{*}$, similarly to their electron counterparts near $B=0$ \cite{Jain.2007, Halperin.PRB.1993, Willett.PRL.1993, Kang.PRL.1993, Willett.PRL.1999, Smet.PRL.1999, Kamburov.PRL.2014}. This phenomenon enables us to directly probe the CF Fermi sea via direct measurements of CF Fermi wavevector. We use a geometric resonance (GR) technique on very high mobility 2DESs, confined to modulation-doped GaAs/AlGaAs heterostructures, and with an imposed, small, periodic density modulation (Fig. 2(a); see Supplemental Materials (SM) \cite{SM} for details). The working principle of GR is straightforward and requires no fitting parameters. The CFs' cyclotron orbit in a small $B^*$ has radius $R_\mathrm{c}^{*} = \hbar k_\mathrm{F}^{*}/eB^{*}$, the size of which is determined by the magnitude of the CFs' Fermi wavevector, $k_\mathrm{F}^{*}$ \cite{ Jain.2007, Halperin.PRB.1993, Willett.PRL.1993, Kang.PRL.1993, Willett.PRL.1999, Smet.PRL.1999, Kamburov.PRL.2014}. If the CFs have a long mean-free-path so they can complete a cyclotron orbit without scattering, then a GR occurs when the orbit diameter becomes commensurate with the period ($a$) of the density modulation [Fig. 2(a)]. Quantitatively, it is generally assumed that, when $2R_\mathrm{c}^{*}/a=i+1/4$ ($i=1,2,3,...$), GRs manifest as minima in magneto-resistance at \cite{ Willett.PRL.1999, Smet.PRL.1999, Kamburov.PRL.2014}:
\begin{equation} \label{Eq1}
 B^{*}_i=2\hbar k_\mathrm{F}^{*}/ea(i+1/4). 
 \end{equation}
 Thus, $k_\mathrm{F}^{*}$ can be deduced directly from the positions of $B^{*}_i$. 

Figure 2(b) highlights our representative magneto-resistance traces over a wide range of $n$, each exhibiting well-developed GR features (marked by arrows), flanking a deep, V-shaped minimum at $\nu = 1/2$. The traces attest to the high sample quality as evidenced by the emergence of fractional quantum Hall states, such as those at $\nu = 1/3$ and $2/3$, even at very low $n$.  In Fig. 2(c), we zoom in close to $\nu = 1/2$. There is a pronounced \textit{asymmetry} in the experimental GR data with respect to the field position of $\nu=1/2$: $|B^*|$ for the GR minimum on the $B^*>0$ side is larger than on the $B^*<0$ side. Figure 2(d) shows the same data as in Fig. 2(c), now plotted as a function of $\nu$. The observed GR minima positions are also asymmetric in $\nu$ positions with respect to $\nu=1/2$. The observed asymmetries are consistent with the data of Kamburov $et$ $al.$ \cite{Kamburov.PRL.2014} which were taken at higher $n$ ($> 12 \times 10^{10}$ cm$^{-2}$).

The data of Figs. 2(c) allow us to directly measure, from the magnetic field positions of the observed GR minima and using Eq. (1) (with $i=1$), the CFs' Fermi wavevector over a large density range, as shown in Fig. 3. We can then address certain fundamental questions: What determines the CFs’ density, Fermi wavevector, and Fermi sea area? What are the implications for the Luttinger theorem and particle-hole symmetry near $\nu=1/2$? These questions have triggered enormous theoretical interest \cite {Barkeshli.PRB.2015, Kachru.PRB.2015, Son.PRX.2015, Senthil.2015, Balram.PRL.2015, Balram.PRB.2016, Metlitskil.2016, Wang2.PRB.2016, Wang3.PRB.2016, Mulligan.PRB.2016, Geraedts.Science.2016, Zucker.PRL.2016, Balram.PRB.2017, Wang.PRX.2017, Cheung.PRB.2017, Pan.NatPhys.2017, Geraedts.PRL.2018, Goldman.PRB.2018, Son.Annul.Rev.Cond.Mat.Phys.2018, Mitra.PRB.2019}. For the first question, three plausible answers are \cite{footnote.spin.polarization} [see Fig. 1 (b)]: $(i)$ It is determined simply by the density of electrons $n$, i.e., the density of CFs ($n^*$) is fixed and equals $n$, and thereby $k_\mathrm{F}^* = (4\pi n)^{1/2}$. $(ii)$ It is the \textit{minority} carriers in the lowest LL that determine $k_{\mathrm{F}}^{*}$, namely $n^* = n_\mathrm{min}$; this means $k_\mathrm{F}^{*}=(4 \pi n_\mathrm{min})^{1/2}$, where $n_\mathrm{min} = n$ for $\nu<1/2$, and $n_\mathrm{min}= n(1-\nu)/\nu$ (i.e., the density of holes in the lowest LL) for $\nu>1/2$. This was the conclusion reached in experiments of Ref. \cite{Kamburov.PRL.2014} and in the subsequent numerical calculations of Refs. \cite{Balram.PRL.2015, Balram.PRB.2016}. $(iii)$ It is equal to half the number of flux quanta penetrating the sample, i.e., $n^* = B/(2h/e)$, if the CFs are Dirac fermions \cite{Son.PRX.2015, Geraedts.Science.2016, Wang.PRX.2017, Cheung.PRB.2017, Son.Annul.Rev.Cond.Mat.Phys.2018, Mitra.PRB.2019}. This Dirac CF theory predicts a change in CF density with magnetic field and renders $k_\mathrm{F}^*= l_\mathrm{B}^{-1} = (4\pi n)^{1/2}\times (B_{i=1}/B_{\nu=1/2})^{1/2}$, where $l_\mathrm{B}=(\hbar/eB)^{1/2}$ is the magnetic length \cite{footnote.HLR}. In addition to the built-in particle-hole symmetry\cite{Son.PRX.2015}, the Dirac CF theory entails a single Dirac cone, as opposed to multiple Dirac cones that are present in 2DESs such as graphene, and might point to a deep relationship between the CF liquid and the three-dimensional topological insulators \cite{Senthil.2015, Metlitskil.2016, Wang2.PRB.2016}. Besides testing the validity of the Luttinger theorem and particle-hole symmetry, one goal of our work is to differentiate between the three possibilities \textit{(i)}-\textit{(iii)} using the new experimental data.


\begin{figure*}[t!]
\includegraphics[width=0.99\textwidth]{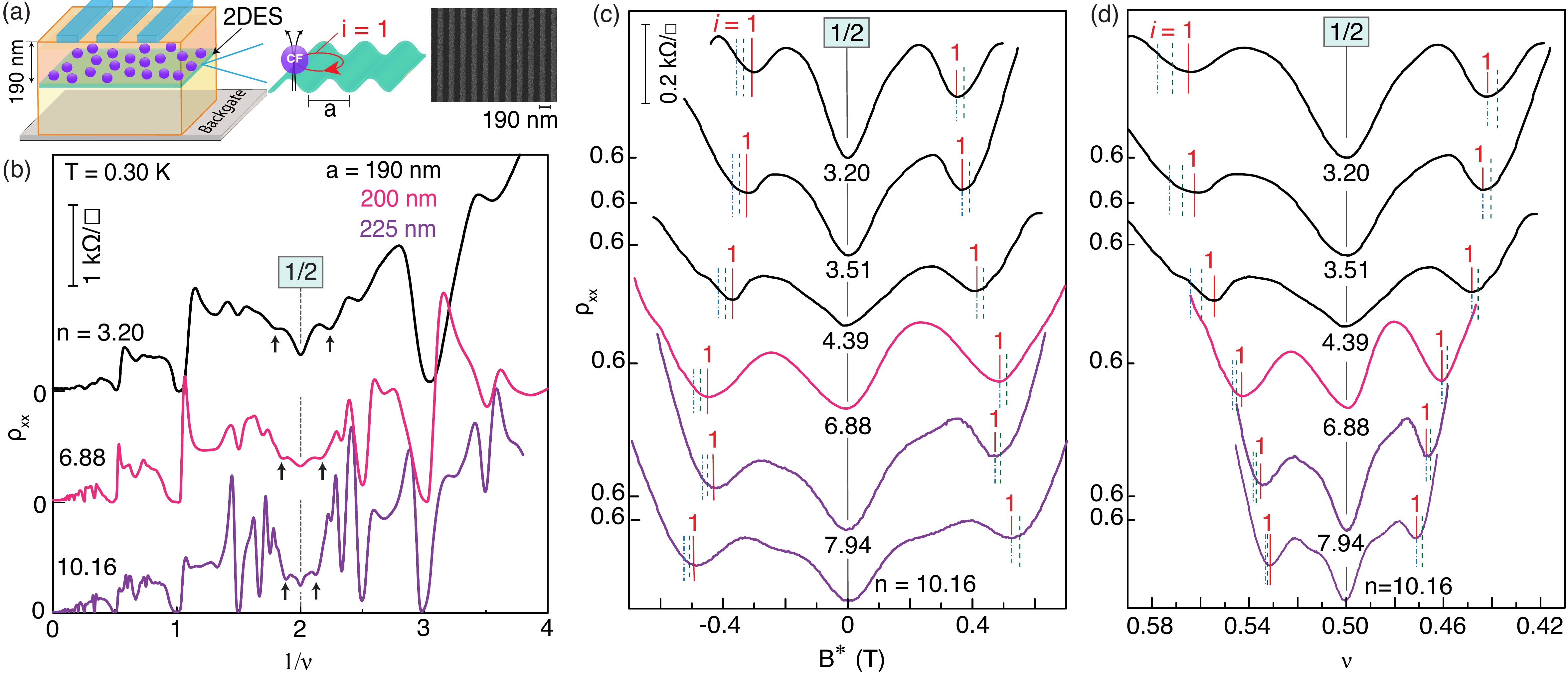}
\caption{\label{fig:Fig2} Overview of our GR technique and magneto-transport data. (a) Our experimental technique consists of patterning a one-dimensional superlattice (shown in blue) on the sample surface to induce a small, periodic density perturbation of period $a$ in the 2DES. A representative scanning electron micrograph shown on the right attests to the uniformity of the stripes. When the cyclotron orbit of the CFs becomes commensurate with $a$, the $i = 1$ GR occurs. (b) Magneto-resistance traces over a wide range of 2DES densities $n$, taken at $T=0.30$ K, plotted against $1/\nu$, showing pronounced GR resistance minima on the flanks of $\nu=1/2$ (vertical arrows), even at very low $n$. The values of $n$ (in units of $10^{10}$ cm$^{-2}$) are given for each trace. (c-d) Expanded view of CF GR features, plotted against $B^*$ and $\nu$. The observed GR minima positions exhibit clear asymmetry with respect to $\nu=1/2$ ($B^* = 0$). Vertical (dash-dotted) blue, (solid) red, and (dashed) green lines mark the $expected$ positions for the $i = 1$ GR for fully spin-polarized CFs according to the fixed density model, minority-carrier model, and Dirac theory, respectively; see text for a description of the models. The blue lines in (c) are exactly symmetric in their positions with respect to $B^*=0$. Also, the blue and red lines coincide for $\nu<1/2$ ($B^*>0$). The experimental data best match the predictions of the minority-carrier model (red vertical lines). The differences between the observed minima positions and the predictions of the fixed density model and Dirac theory are also clearly visible.}
\end{figure*}  

\begin{figure}[t!]
\includegraphics[width=.48\textwidth]{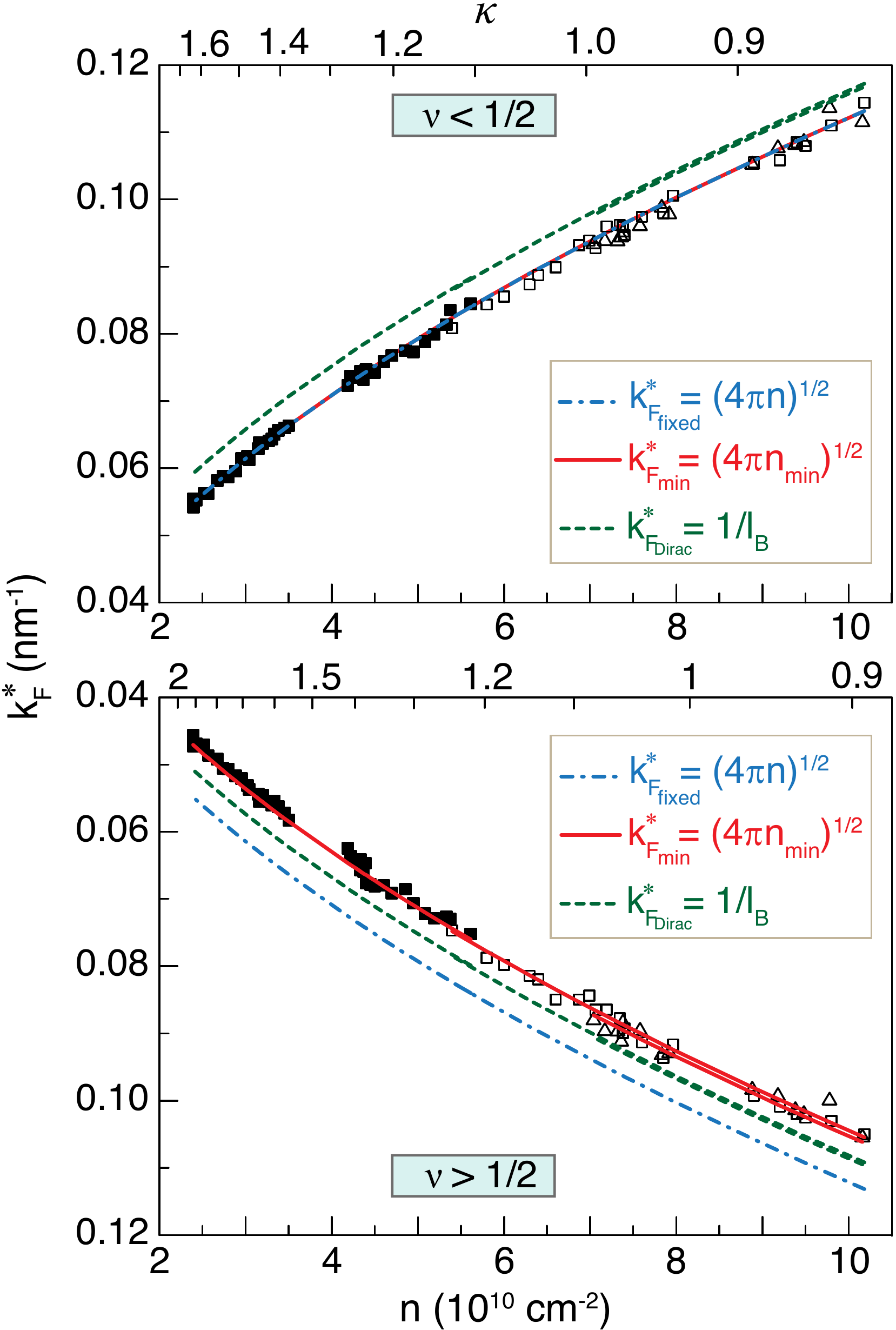}
\caption{\label{fig:Fig3} 
CF Fermi wavevectors determined from the measured GR minima plotted against density $n$. The symbols represent experimental data from samples with modulation periods $a=190$ nm (filled squares), $a=200$ nm (open squares), and $225$ nm (open triangles), respectively. Blue, red, and green curves represent the calculated $k_\mathrm{F}^*$ based on $k_\mathrm{F}^* = (4\pi n)^{1/2}$, $k_\mathrm{F}^* = (4\pi n_\mathrm{min})^{1/2}$, and Dirac theory, respectively. For each model, the results of calculations are shown in different ranges of $a$ where the experimental data were taken. For a description of the multiple curves for the Dirac theory and minority-carrier model, see \cite{multiple}. The experimental data match the minority-carrier expression (red curves) very well. The top axes give the LL mixing parameter $\kappa$.}
\end{figure} 

\begin{figure}[t!]
\includegraphics[width=.49\textwidth]{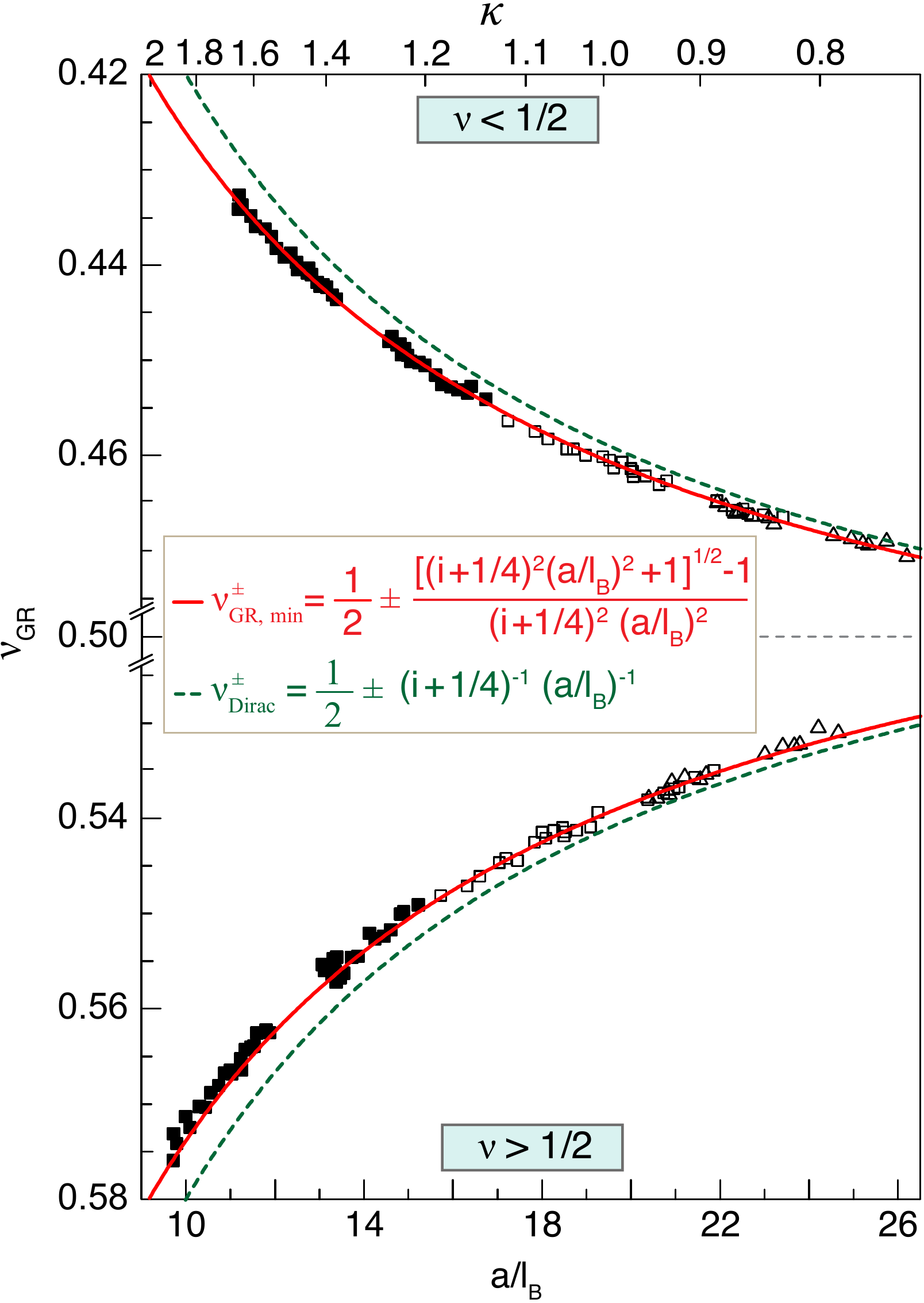}
\caption{\label{fig:Fig4} The filling-factor positions of the observed CF GR minima $\nu_\mathrm{GR}$ at different densities, plotted against $a/l_B$. The red and green curves are the predictions of the minority-carrier model and Dirac theory, respectively (expressions are given in the inset).}
\end{figure}

First, we address the question of what determines the CF Fermi sea area and whether this area is independent of the inter-CF interaction, just as Luttinger predicted. In Fig. 3 we show the CFs' $k_\mathrm{F}^*$ predicted by the models $(i)$-$(iii)$. It is clear in Fig. 3 that the \textit{minority-carrier} model best fits the experimental data throughout our density range. This is similar to the single-particle picture, e.g., in doped semiconductors: In $n$-doped systems, the area of the Fermi sea is determined by the density of electrons in the conduction band while in $p$-doped systems, where the valence band is almost full, the Fermi sea area is determined by the density of the empty states in the valence band, i.e., ``holes," rather than electrons. Remarkably, $k_\mathrm{F}^*$ and therefore the area of the CF Fermi sea follows the same simple rule.

Furthermore, the Luttinger theorem postulates that the Fermi sea area should be independent of the interaction between the fermions. Our low-density data provide very important verification here. Even though the non-perturbative part of the electron-electron interaction is already used in making the CFs, the residual interaction between the CFs increases substantially in the low-density regime thanks to the increased mixing between the LLs \cite{Jain.2007, Zhang.PRL.2016}. This effect can be quantified by the LL mixing (LLM) parameter $\kappa = E_\mathrm{Coul}/E_\mathrm{Cyc}$, defined as the ratio of the Coulomb energy ($E_\mathrm{Coul} = e^2/4 \pi \varepsilon l_\mathrm{B}$) to the cyclotron energy ($E_\mathrm{Cyc}=\hbar eB/m^*$). Note that $\kappa$, whose values are indicated on the top axes of Fig. 3 plots, goes as $n^{-1/2}$ at a fixed $\nu$. The LLM and the resulting interaction between the CFs affect the CF ground state significantly. They can lead, e.g., to a transition to a CF Wigner crystal \cite{Zhao.PRL.2018}. However, the data of Fig. 3 show that $k_\mathrm{F}^*$,  which determines the area of the CF Fermi sea, follows the same expression $k_\mathrm{F}^* = (4\pi n_\mathrm{min})^{1/2}$ over a large range of electron density and LLM, making a convincing case that the Luttinger theorem is obeyed in a strongly interacting system.

Next, we discuss the asymmetries observed in Figs. 2(c,d) and Fig. 3 with respect to $\nu=1/2$. These are puzzling at first sight, and might imply a breakdown of particle-hole symmetry. The question of whether or not the CFs obey the particle-hole symmetry has in fact sparked exciting new developments in the field of strongly interacting electron systems \cite {Barkeshli.PRB.2015, Kachru.PRB.2015, Son.PRX.2015, Senthil.2015, Balram.PRL.2015, Balram.PRB.2016, Metlitskil.2016, Wang2.PRB.2016, Wang3.PRB.2016, Mulligan.PRB.2016, Geraedts.Science.2016, Zucker.PRL.2016, Balram.PRB.2017, Wang.PRX.2017, Cheung.PRB.2017, Pan.NatPhys.2017, Geraedts.PRL.2018, Goldman.PRB.2018, Son.Annul.Rev.Cond.Mat.Phys.2018, Mitra.PRB.2019, Sreejith.PRB.2017}. These theories mostly predict that the particle-hole symmetry should hold within the lowest LL, while Balram \textit{et al.}, \cite{Balram.PRL.2015} conclude that it could be broken when LLM is significant. 

In theory, particle-hole symmetry about $\nu=1/2$ implies the equivalency of $\nu \leftrightarrow (1-\nu)$ at a fixed $B$. However, in our experiments where we vary $B$ while keeping $n$ fixed, the $i=1$ GR minima for $\nu>1/2$ and $\nu<1/2$ are observed at two different absolute values of $B^\pm$ (and therefore $l_\mathrm{B}$). As a result, for a sample with a fixed $a$ and $n$, the relevant parameter $a/l_\mathrm{B}$, is different at $\nu$ and $(1-\nu)$. This can be accounted for by plotting the data as a function of $a/l_B$. Figure 4 illustrates such plots where $\nu$ at which the GR occurs ($\nu_\mathrm{GR}$), taken directly from the experimental traces, are shown against $a/l_\mathrm{B}$ (we show $B_{i=1}^*$ and $k_\mathrm{F}^*$  vs. $a/l_\mathrm{B}$ plots in the SM \cite{SM}). Remarkably, the experimental GR data, when plotted in this fashion, are \textit{symmetric} with respect to $\nu=1/2$ in the entire density range, within the experimental accuracy. This leads us to a very important conclusion: The GR data \textit{are} consistent with particle-hole symmetry about $\nu=1/2$, even at small $n$ (small $a/l_\mathrm{B}$) where the LLM and the inter-CF interaction are significant. The asymmetry in $\nu_\mathrm{GR}$ with respect to $\nu=1/2$ [Fig. 2(d)] does not imply that particle-hole symmetry is broken; the apparent asymmetry emerges only because, in a given experiment at a fixed density, the parameter $a/l_\mathrm{B}$ is not identical at $\nu$ and $(1-\nu)$.

Interestingly, the particle-hole symmetry in Fig. 4 data can be easily understood from the same expression $k_\mathrm{F}^* = (4\pi n_\mathrm{min})^{1/2}$ that we find to be a good representation of Fig. 3 data. This expression can be written as:
 \begin{equation} \label{Eq2}
   k_\mathrm{F}^* = \left\{
   \begin{aligned}
     (2\nu)^{1/2} l_\mathrm{B}^{-1}; \: \: \: \: \: \: \: \:  \: \: \: \nu<1/2  \\
     [2(1-\nu)]^{1/2} l_\mathrm{B}^{-1}; \nu>1/2
   \end{aligned}
   \right.
\end{equation}
The above expression clearly obeys the particle-hole symmetry $\nu \leftrightarrow (1-\nu)$ about $\nu=1/2$, provided that $l_\mathrm{B}$ is fixed. Using Eqs. (1) and (2), we then work out a quadratic equation for $\nu_\mathrm{GR}$; details are given in SM \cite{SM}:
\begin{equation} \label{Eq3}
\frac{(2\nu_\mathrm{GR, min})^{1/2}}{1-2\nu_\mathrm{GR, min}} =\frac{a}{2l_\mathrm{B}} (i+\frac{1}{4}). 
\end{equation}
Solving this equation, we arrive to a hitherto unknown expression for $\nu_\mathrm{GR, min}$:
\begin{equation} \label{Eq4}
\nu^{\pm}_\mathrm{GR, min} = \frac{1}{2} \pm \frac{\bigg[\big(i+\frac{1}{4}\big)^2 (\frac{a}{l_\mathrm{B}}\big)^2 +1\bigg]^{1/2} -1}{\big(i+\frac{1}{4}\big)^2 \big(\frac{a}{l_\mathrm{B}}\big)^2}. 
\end{equation}

It is clear from Eq. (4) that at a fixed $a/l_\mathrm{B}$, $\nu^{\pm}_\mathrm{GR, min}$ is symmetric about $\nu=1/2$. In Fig. 4 we plot the predictions of the Eq. (4) (red curves) with $i=1$. Similar to Fig. 3, the data of Fig. 4 show excellent agreement with the minority-carrier model. In Fig. 4 we also plot the predictions of the Dirac theory (green curves, see SM \cite{SM})\cite{footnote.HLR}. The Dirac theory also exhibits particle-hole symmetry but it does not agree with the data quantitatively. The low-density (small $a/l_\mathrm{B}$) data and their agreement with the minority-carrier model are again particularly important as they clearly differentiate this model from the Dirac theory.

To place our results in a broader perspective, we compare the CF Fermi sea with other strongly interacting systems whose physics is not well understood \cite{Sachdev.2018}. For example, in high-$T_\mathrm{c}$ cuprate superconductors, the volume of the Fermi surface for large hole dopings is determined by the majority carriers \cite{Hussey.2003, plate.2005}, exactly opposite to the CF Fermi sea. Strangely, however, at low hole dopings, the volume of the Fermi surface appears to be very small and equal to the doping density \cite{Taillefer.2007}. Such subtlety in the Fermi surface remains a mystery. In contrast, we find that the Fermi sea of strongly interacting CFs is always governed by the minority carrier density, Luttinger theorem, and particle-hole symmetry. This suggests that the intricate physics of other Fermi seas harboring strongly interacting electrons could perhaps be simplified by finding the emergent particles of the systems, similar to the CFs in the half-filled LL.

We conclude by making three remarks. First, our direct measurements of CF Fermi wavevector provide quantitative evidence that the Luttinger theorem and particle-hole symmetry are obeyed in a system where the quasi-particles themselves are a product of strong interaction. Second, unlike the cuprates and the heavy fermion compounds such as YbRh$_2$Si$_2$ \cite{Rourke.2008} and EuRh$_2$Si$_2$ \cite{Guttler.2019}, the CF Fermi wavevector and Fermi sea area appear to be determined by the minority-carrier density in the lowest LL. Third, our data show deviations from both Dirac \cite{Son.PRX.2015} and Halperin-Lee-Read \cite{Halperin.PRB.1993} theories \cite{Cheung.PRB.2017, footnote.HLR}. The deviations can possibly be reconciled if one incorporates subtle corrections in the Dirac/Halperin-Lee-Read framework \cite{Mitra.PRB.2019}. This brings up a crucial question. Is it simply a fortuitous coincidence that the results of such sophisticated theories lead to the simple experimental conclusion, namely $k_\mathrm{F}^* = (4\pi n_\mathrm{min})^{1/2}$, even in the strongly interacting limit?

\begin{acknowledgments}
We acknowledge support through the the National Science Foundation (Grant No. DMR 1709076) for measurements, and the U.S. Department of Energy Basic Energy Science (Grant No. DEFG02-00-ER45841), the National Science Foundation (Grants No. ECCS 1906253 and No. MRSEC DMR 1420541), and the Gordon and Betty Moore Foundation’s EPiQS Initiative (Grant No. GBMF9615) for sample fabrication and characterization. M.S. acknowledges a QuantEmX travel grant from the Institute for Complex Adaptive Matter (ICAM) and the Gordon and Betty Moore Foundation through Grant No. GBMF5305. We also acknowledge illuminating discussions with B. I. Halperin and M. Mulligan, and particularly thank  A. C. Balram and J. K. Jain for many discussions and their suggestion for making Fig. 4 plot to elucidate the particle-hole symmetry implied by our data.
\end{acknowledgments}


\begin{thebibliography}{99}
\bibitem{footnote.spin.polarization} The expressions in $(i)$ to $(iii)$ for $k_\mathrm{F}^*$ assume that the CFs are fully spin polarized; for spin-singlet CFs, the factor of 4 in the expression $(4 \pi n)^{1/2}$ should be replaced by 2, implying a $\sqrt{2}$ times smaller $k_\mathrm{F}^*$.
\bibitem{multiple} In Fig. 3 we plot multiple curves for the predictions of Dirac theory (green curves) for both $\nu>1/2$ and $\nu<1/2$, and for the predictions of minority-carrier model (red curves) for $\nu>1/2$. These curves represent the theoretical predictions for the three different superlattice periods, $a=190$ (filled squares), 200 (open squares), and 225 nm (open triangles), that we use in our experiments. We require multiple curves, because the locations of the GR features ($B^*$) and therefore the corresponding $k_\mathrm{F}^*=1/l_\mathrm{B}$ in the Dirac theory depend on the superlattice period $a$. Consequently, the predicted $k_\mathrm{F}^*$ values are different for the three periods, and require three curves to show these predictions. The minority-carrier model on the $\nu>1/2$ side also predicts $k_\mathrm{F}^*$ that depends on $B^*$ and $a$, thus requiring three curves to show the predictions for the three different periods.
\end{thebibliography}
\end{document}